\magnification=\magstep1
\hsize=13cm
\vsize=20cm
\overfullrule 0pt
\baselineskip=13pt plus1pt minus1pt
\lineskip=3.5pt plus1pt minus1pt
\lineskiplimit=3.5pt
\parskip=4pt plus1pt minus4pt

\def\negenspace{\kern-1.1em}



\newcount\secno
\secno=0
\newcount\susecno
\newcount\fmno\def\z{\global\advance\fmno by 1 \the\secno.
                       \the\susecno.\the\fmno}
\def\section#1{\global\advance\secno by 1
                \susecno=0 \fmno=0
                \centerline{\bf \the\secno. #1}\par}
\def\subsection#1{\medbreak\global\advance\susecno by 1
                  \fmno=0
       \noindent{\the\secno.\the\susecno. {\it #1}}\noindent}


\def\sqr#1#2{{\vcenter{\hrule height.#2pt\hbox{\vrule width.#2pt
height#1pt \kern#1pt \vrule width.#2pt}\hrule height.#2pt}}}
\def\square{\mathchoice\sqr64\sqr64\sqr{4.2}3\sqr{3.0}3}




\def\neq{\hbox{$\,$=\kern-6.5pt /$\,$}}





\newcount\secno
\secno=0
\newcount\fmno\def\z{\global\advance\fmno by 1 \the\secno.
                       \the\fmno}
\def\sectio#1{\medbreak\global\advance\secno by 1
                  \fmno=0
       \noindent{\the\secno. {\it #1}}\noindent}


\def\kg{[\![}
\def\gk{]\!]}

\def\la{\langle}
\def\ra{\rangle}




\magnification=\magstep1
\hsize 13cm
\vsize 20cm

\vglue 2.0cm
 \centerline{\bf{HIGHER-DERIVATIVE
BOSON FIELD THEORIES}} \centerline{\bf{AND CONSTRAINED
SECOND-ORDER THEORIES}}

\vskip 0.3cm
\centerline{by}
\vskip 0.7cm
\centerline{
                F.J. de Urries$^{(*)(\dagger)}$,
                J.Julve$^{(\dagger)}$ and
                E.J.S\'anchez$^{(\dagger)(\diamond)}$
}

\vskip 0.2cm \centerline{ $^{(*)}$\it Departamento de F\'\i sica,
Universidad de Alcal\'a de Henares,} \centerline{\it 28871
Alcal\'a de Henares (Madrid), Spain} \vskip 0.2cm \centerline{
$^{(\dagger)}$\it{IMAFF, Consejo Superior de Investigaciones
Cient\'\i ficas,}} \centerline{\it{Serrano 113 bis, Madrid 28006,
Spain}} \vskip 0.2cm \centerline{
        $^{(\diamond)}${\it Departamento de Matem\'atica,
        Universidad Europea,} }
\centerline{\it 28670 Villaviciosa de Od\'on (Madrid), Spain}

\vskip 0.7cm

\centerline{\bf ABSTRACT}\bigskip As an alternative to the
covariant Ostrogradski method, we show that higher-derivative
relativistic Lagrangian field theories can be reduced to second
differential-order by writing them directly as covariant
two-derivative theories involving Lagrange multipliers and new
fields. Despite the intrinsic non-covariance of the Dirac's
procedure used to deal with the constraints, the explicit Lorentz
invariance is recovered at the end. We develop this new setting
on the grounds of a simple scalar model and then its applications to
generalized electrodynamics and higher-derivative gravity are worked out.
For a wide class of field theories this method is better suited than
Ostrogradski's for a generalization to $2n$-derivative theories.
\bigskip
\centerline{PACS numbers: 11.10.Ef, 11.10.Lm, 04.60}

\vfill

\noindent{\it e-mail addresses: julve@imaff.cfmac.csic.es;
fernando.urries@alcala.es;}

\quad\quad\quad\quad\quad\quad {\it ejesus.sanchez@mat.ind.uem.es}

\eject


\sectio{\bf{Introduction}}\bigskip

Higher-derivative (HD) theories have an old tradition in physics.
Relativistic field theories with higher order Lagrangians
historically range from Higgs model regularizations [1] to
generalized electrodynamics [2][3] and HD gravity [4], and arise
as effective low energy theories of the string [5] or induced by
quantum fields in a curved background [6]. A procedure was later
devised to reduce them, by a Legendre transformation, to
equivalent lower-derivative (LD) second-order theories [7] where a
subsequent diagonalization explicitly displays the particle
degrees of freedom [3][8][9].

The validity of the formal Lorentz covariant order-reducing method
adopted there has been checked in an example of scalar HD theories
by a rigorous study of the phase-space [9]. In this procedure, a
generalization of the Ostrogradski formalism to continuous
relativistic bosonic systems ($2n$-derivative because of Lorentz
invariance in the most usual cases) is carried out. In it, some of
the field derivatives and the generalized conjugate momenta
become, after a suitable diagonalization, new field coordinates
describing the degrees of freedom (DOF) which were already
identified in the particle propagators arising in the algebraic
decomposition of the HD propagator.

By using Lagrange multipliers, an alternative to the Ostrogradski
method for mechanical discrete systems has been proposed which
allows to show the quantum (Path Integral) equivalence between the
modified action principle (first order Helmholtz Lagrangian) and
the starting HD theory [10]. For relativistic field systems, we
show that a similarly inspired procedure can be followed in which
the multipliers let to write the HD theory from the outset as a
second order (constrained) covariant one which lends itself to a
particle interpretation after diagonalization.

In this paper we implement this new setting by means of the use of
Lagrange multipliers in a Lorentz invariant formulation of
relativistic scalar (and subsequent applications to vector and
tensor) field theories. In the following we will generically refer
to them as HD theories (HDT). The Dirac method [11] prescribes the
identification of the primary constraints arising in the
definition of the momenta. These constraints are added to the
starting Hamiltonian by means of new multipliers, and then they
are required to be conserved by the time evolution driven by this
enlarged ``total Hamiltonian" through the Poisson Bracket (PB).
This may give rise to secondary constraints, the conservation of
which can in turn generate more secondary constraints. The process
stops when the equation obtained is not a constraint but an
equation allowing the determination of a  multiplier. We then use
the remaining constraints to eliminate the remaining multipliers
and the momenta, ending up with a two-derivative theory depending
on just its true DOF. Since the latter appear mixed, a
diagonalization works finally out the independently propagating
DOF.

As long as time evolution is analyzed, the true mechanical
Hamiltonian (i.e. the energy) of the system must be used. Then one
cannot benefit of the compactness of the Lorentz invariant
procedures introduced in [9], so we are initially forced to deal
with non covariant 3+1 objects and face the diagonalization of
larger matrices. The relativistic invariance of the system becomes
explicit only at the end of the process. Alternatively this
invariance may also be maintained at an earlier stage by merging
the constraints into covariant equations: in fact it is possible
to find all the previously known 2nd class (non-covariant) Dirac's
constraints by working out the whole set of covariant field
equations (for fields and multipliers).

From the methodological point of view, the new treatment of HD
boson theories that we present here provides a sharp departure
from the more traditional Ostrogradski approach. Moreover, it is
implementable and may prove advantageous in generalized
electrodynamics, HD Yang-Mills and linearized HD gravity as well.
On the other hand it lends itself to generalization to arbitrary
$n$ better than the Ostrogradski method does.

Our work focuses on the propagating DOF of these theories.
Therefore we mainly consider the free part of the corresponding
(HD and their equivalent LD) Lagrangians. Self interactions
(derivative or not) and interactions with other (external) fields
are embodied in a source term, namely a source $j$ linearly
coupled to the fundamental field. The source always contains a
coupling constant. Thus for our purposes one may adopt two
attitudes: either $j$ is assumed to contain only (spectator)
external fields or the whole source term is treated as a
perturbation (and ignored at the zeroth order in the coupling we
are interested in) if self couplings are present. In any case we
retain an external source as a guide to find the suitable
diagonalizing matrix.

HD (free) fermion theories will always be $(2n+1)$-derivative to
maintain Lorentz invariance. The constraint method presented here
(as well as Ostrogradski's) can be adapted to encompass these
theories, but the task is beyond the purposes of this paper.

Again resorting to the scalar example, in Section 2 we review the
general situation of HD theories in which the ``square masses" may
be the complex and/or degenerate roots of a polynomial, then
justifying the custom of focusing only on theories with real
non-degenerate squared masses. In Section 3 we treat $n=3$, the
case $n=2$ being too much trivial for our illustrative purposes.
The results regarding the extension to arbitrary $n$ are presented
in Section 4. Through the Dirac's canonical method followed, they
guarantee that all the constraints present in the theory have been
properly taken into account. This validates the Lorentz covariant
phase-space method described in Section 5 which we adopt in the
following as the suitable shortcut for practical uses.
 Section 6 discusses,
with the appropriate mathematical formalism, the application of
our constraint method to HD (Podolsky) vector field theories and
generalizations together the most interesting case of HD (linearized)
gravity. We summarize our results in the Conclusions in
Section 7.

In Appendix A we show the pure algebraic character (i.e. absence of space
derivatives) of the secondary constraints.  Also for the general
 $2n$-derivative case, in Appendices B and C we display the steps
leading to the diagonalization of the DOF.
In Appendix D, when squares of the multipliers occur,
we show the equivalence between this seemingly constraint method and
Ostrogradki's.

Throughout the paper
the Minkowski metric $(+ - - -)$ is adopted.

\vfill
\eject


\sectio{{\bf HD Scalar Theories}}

        The theories of arbitrary (finite) differential order have a distinctive
spectrum with respect to the usual second order theories
describing physical systems. It generally includes peculiar
ghostly states and non-particle solutions. Although this feature
is not exclusive of HDT's (one may devise second order theories
with this sort of complications), it is almost inherent to them.
The exception corresponds to very special cases like HD gravity
made up only with powers of the scalar curvature. There is no
escape however for simpler fields.

The simplest HDT's are linear theories of scalar fields
$${\cal L}^{ ^{N} }=-{c_{_{N}}\over 2}\phi q_{_{N}}\!(\square) \phi -j\phi\quad,
\eqno(\z)
$$
\noindent{where} $c_{_{N}}$ is a constant with suitable dimensions
and $q_{_{N}}( \square)$ is a monic real polynomial in $
\square\,$. A detailed study of the theories (2.1) in which
$q_{_{N}}$ has only real non-degenerate roots can be found in [9].
In these cases, the theory (2.1) is equivalent to a LD one, made
up of the alternate sum of $N$ Klein-Gordon (KG) free Lagrangians
where the sum of the respective real fields couples to the source
$j$. This states that all the DOF are particle-like and couple to
the same source.

However this is a privileged case of (2.1) as long as $q_{_{N}}$ may
generally have degenerate and/or complex roots
$$
\eqalign{q_{_{N}}\!( \square)=&( \square+m_{1}^{2})^{R_{1}}
        \cdots( \square+m_{r}^{2})^{R_{r}}
                ( \square-M_{1}^{2})^{T_{1}}\cdots( \square-M_{1}^{2})^{T_{t}}\cr
&[( \square-\Omega_{1})( \square-\bar{\Omega}_{1})]^{C_{1}}\cdots
[( \square-\Omega_{c})( \square-\bar{\Omega}_{c})]^{C_{c}}}
$$
\noindent{where} the $m_i$ correspond to physical masses, $M_i$
are tachionic and $\Omega_i$ are complex, $R_i$, $T_i$ and $C_i$
being their respective degeneracies.

Non degenerate complex roots can be handled formally with the
methods developed in [9]. Following them, it can be seen that
$$
{\cal L}^{4}=-{c_{4}\over 2}\phi( \square-\Omega)(
\square-\bar{\Omega})\phi-j\phi
$$
\noindent{is} equivalent to
$$
{\cal L}^2={{c_{4}}\over 2}(\Omega-\bar{\Omega})\left[\varphi(
\square-\Omega) \varphi -\bar{\varphi} ( \square-\bar{\Omega})
\bar{\varphi} \right] -j(\varphi+\bar{\varphi})\quad, \eqno(\z)
$$
\noindent{where} $\varphi=\varphi_{1}+i\varphi_{2}$ is a complex
scalar field. However this is not a real diagonalization because
(2.2) cannot be diagonalised in terms of real (or even complex)
scalar fields built linearly out of $\varphi_{1}$ and
$\varphi_{2}$ and with real square masses.

Also the degeneracy makes it impossible to interpret the roots as
the square masses of free particle-like (even tachionic) states.
In fact, already the simple case  $\phi (\square+m^2)^2\phi$
cannot be reduced to a LD theory of free real (or complex) fields
because the HD propagator  $1/(\square +m^2)^2$ is not a linear
combination of LD free propagators $1/( \square \pm {\rm
m}^2_i)\,$. Otherwise stated, there are solutions to $(\square
+m^2)^2\phi=0$ that are not solutions to $(\square +m^2)\phi=0$,
and hence cannot be expressed as superpositions of plane waves.

Summarizing, only in the case of real non-degenerate square masses
can the HD propagator be algebraically expanded as a sum of
particle-like propagators, i.e. in terms of propagators displaying
physical or tachionic masses (sign of the square mass), or even a
non-physical norm (sign of the propagator). As customary in the
literature, in the following we will limit ourselves to the study
of these cases.


\sectio{\bf n = 3 theory}
\bigskip

We consider the six-derivative Lagrangian,

$${\cal L}^{6} = -{1 \over 2}{\mu ^2 \over {M}}
              \,\phi \kg 1\gk\kg 2\gk\kg 3\gk \phi
                                          -j\,\phi\quad ,\eqno(\z)$$

\noindent{where} $\mu$ is an arbitrary mass parameter, $\kg i\gk
\equiv (\square + m^2_i)$  are KG operators, $M \equiv \langle 12
\rangle \langle 13 \rangle \langle 23
\rangle$\enskip,\enskip$\langle ij \rangle \equiv m_i^2 -
m_j^2>0$\enskip for\enskip$i < j $\enskip, and mass dimensions
$[\mu ]=1$, $[M]=6$, $[\phi]=1$, $[j]=3$. We have overlooked the
case $n=2$ because it is too trivial for introductory purposes.

As discussed in [9], (3.1) displays the general form of the free
part of a higher-derivative scalar theory with non-degenerate
masses $m_1\,$, $m_2\,$, $m_3\,$, the source term embodying the
remaining self-interactions and the couplings to other fields.
There we showed by a covariant Legendre order-reduction procedure
that (3.1) is equivalent to the second-order Lagrangian

$$ {\cal L}^{2} = -{1 \over 2}{\mu ^2 \over {\langle 23\rangle}}
                                                \,\phi_1 \kg 1\gk\phi_1
                 +{1 \over 2}{\mu ^2 \over {\langle 13\rangle}}
                                                \,\phi_2 \kg 2\gk\phi_2
                 -{1 \over 2}{\mu ^2 \over {\langle 12\rangle}}
                                                \,\phi_3 \kg 3\gk\phi_3
                 -j\,(\phi_1+\phi_2+\phi_3)\quad ,\eqno(\z)$$

\noindent{in} agreement with what is expected from the algebraic expansion of the HD propagator
in particle poles, namely

$$-{{\mu ^{-2}M} \over {\kg 1\gk\kg 2\gk\kg 3\gk}}
       = -{{\mu ^{-2}\langle 23\rangle} \over {\kg 1\gk}}
         +{{\mu ^{-2}\langle 13\rangle} \over {\kg 2\gk}}
         -{{\mu ^{-2}\langle 12\rangle} \over {\kg 3\gk}}\quad .\eqno(\z)$$

\noindent{The} physical meaning is that whenever a source $j$ emits a field
$\phi$ driven by a HD dynamics, it actually emits a linear superposition
$\phi=\phi_1+\phi_2+\phi_3$ of LD (particle) fields.

However ${\cal L}^{6}$ can be recast directly as a second-order
theory with constraints, namely

$${\cal L}^{6}={1\over 2}{\mu^2 \over M}\bigl[-{\bar\psi}_3 \kg 1\gk
{\bar\psi}_1 + \lambda_1 ({\bar\psi}_1 -\kg 2\gk {\bar\psi}_2)
+\lambda_2 ({\bar\psi}_2 -\kg 3 \gk{\bar\psi}_3)\bigl]
-j{\bar\psi}_3\quad, \eqno(\z)$$

\noindent{where} ${\bar\psi}_3=\phi$ and $\lambda_1\,$,
$\lambda_2$ are Lagrange multipliers, so that ${\cal L}^{6}$
depends on five fields. It is immediate to check that the
equations of motion for $\lambda_1$ and $\lambda_2$ yield
${\bar\psi}_1$ and ${\bar\psi}_2$ as (derivative) functions of
${\bar\psi}_3$. By substituting them in (3.4) and identifying
${\bar\psi}_3$ and $\phi$ one recovers the HD equation (3.1).
Dropping a total time-derivative, in compact matrix notation,
(3.4) reads

$${\cal L}^{6}={1\over 2}{\dot\Psi^{T}{\cal K}\,{\dot\Psi}}+{1\over
2}\Psi^{T}{\cal M}\,\Psi -J^{T}\Psi\quad, \eqno(\z)$$

\noindent{where} the vectors $\Psi$ and $J$, with components
$\psi_i\,$, $J_i\,$, and the matrices ${\cal K}$ and ${\cal M}$
are

$$\Psi\equiv\left(\matrix{\mu^{-4}{\bar\psi}_1\cr\mu^{-
2}{\bar\psi}_2\cr{\bar\psi}_3\cr
\mu^{-2}\lambda_1  \cr\mu^{-4}\lambda_2\cr}\right)\quad
{\rm so}\enspace {\rm that}\quad [\psi_i]=1\quad i=1,\dots ,5\quad
;\quad J_i=j\delta_{3i}\quad ;$$
$${}\eqno(\z)$$
$${\cal K}={\mu^{6}\over 2M}\left(\matrix{0&0&1&0&0\cr
0&0&0&1&0\cr 1&0&0&0&1\cr 0&1&0&0&0\cr 0&0&1&0&0}\right);\;{\cal
M}={\mu^{6}\over 2M}\left(\matrix{0&0&-M_1^2&\mu^{2}&0\cr
0&0&0&-M_2^2&\mu^{2}\cr -M_1^2&0&0&0&-M_3^2\cr
\mu^{2}&-M_2^2&0&0&0\cr
0&\mu^{2}&-M_3^2&0&0}\right). $$

\noindent{$\cal M$} is an operator with space derivatives present
in $M_i^2\equiv m_i^2-\Delta\,$.

The canonical conjugate momenta are defined as
$$\pi_i={\partial {\cal L}^6\over \partial \dot\psi_i}\quad.\eqno(\z)$$
\noindent{They} are the components of a 5-vector $\Pi$ for which
one has

$$\Pi={\cal K}{\dot\Psi}\quad.\eqno(\z)$$

Since ${\cal K}$ is not invertible, not all the velocities
$\dot\psi_i$ can be expressed in terms of the momenta and a
primary constraint occurs, namely

$$\Omega_1\equiv\pi_5-\pi_1=0\quad,\eqno(\z)$$

\noindent{as} consequence of $\pi_5={\mu^6\over
2M}\dot\psi_3=\pi_1\,$. There is only one such a constraint since
the submatrix ${\cal K}_{ab}\equiv{\mu^{6}\over 2M}{\cal
K}_{ab}^{'}\quad (a,b=1,\dots ,4)$ is regular. In the following,
indices $a,b,...$ go from 1 to 4, while $i,j,...$ go from 1 to 5.
The velocity $\dot\psi_5$ is not worked out, and from (3.8) we
have

$$\pi_a={\mu^6\over 2M}{\cal K}_{ab}^{'}\dot\psi_b+{\mu^6\over
2M}\delta_{a3}\dot\psi_5\quad,\eqno(\z)$$

\noindent{and} therefore

$$\dot\psi_a={2M\over\mu^6}{\cal K}_{ab}^{'}\pi_b-
\delta_{a1}\dot\psi_5\quad.\eqno(\z)$$
\vfill
\eject

The Hamiltonian is

$${\cal H}=\pi_a\dot\psi_a+\pi_5\dot\psi_5-{1\over 2}\dot\psi_a{\cal
K}_{ab}\dot\psi_b-{\mu^6\over 2M}\dot\psi_3\dot\psi_5-{1\over
2}\psi_i{\cal M}_{ij}\psi_j+j\psi_3\quad,\eqno(\z)$$

\noindent{where} $\dot\psi_a$ must be substituted according to
(3.11). Then the dependence on $\dot\psi_5$ cancels out and we
have

$${\cal H}={1\over 2}{2M\over \mu^6}\pi_a{\cal K}_{ab}^{'}\pi_b-{1\over
2}\psi_i{\cal M}_{ij}\psi_j+J_i\psi_i\quad.\eqno(\z)$$

Since not all of the five momenta $\pi_i$ are independent because
of the primary constraint (3.9), only four momenta appear in
(3.13) together with the five fields$\;\psi_i\;$. The ``total
Hamiltonian", with five independent momenta, accounting for this
is

$${\cal H}_T={\cal H}+\zeta\Omega_1\quad,\eqno(\z)$$
\noindent{where} $\zeta$ is a Lagrange multiplier.

The stability of $\Omega_1$ requires

$$\dot\Omega_1=\bigl\{\Omega_1,{\cal
H}_T\bigr\}_{PB}\equiv\Omega_2={\mu^6\over 2M}\bigl(\la
13\ra\psi_3- \mu^2\psi_4+\mu^2\psi_2\bigr)=0\quad.\eqno(\z)$$

\noindent{This} secondary constraint yields

$$\psi_4={\la 13\ra\over\mu^2}\psi_3+\psi_2\quad.\eqno(\z)$$                                                                                                                                                                                                                                                                                                                                                                                                                                                                   
Further secondary constraints stem from the ensuing stability conditions

$$\dot\Omega_2=\bigl\{\Omega_2,{\cal H}_T\bigl\}_{PB}\equiv\Omega_3=\la
13\ra\pi_1-\mu^2\pi_2+\mu^2\pi_4=0\quad,\eqno(\z)$$

\noindent{so} that

$$\pi_4=\pi_2-{\la 13\ra\over\mu^2}\pi_1\quad,\eqno(\z)$$

\noindent{and} again

$$\eqalign{\dot\Omega_3&=\bigl\{\Omega_3,{\cal
H}_T\bigr\}_{PB}\equiv\Omega_4=\cr&={\mu^6\over 2M}\bigl(-\la
13\ra\la 23\ra\psi_3-\mu^2\la 13\ra\psi_2-
\mu^4\psi_1+\mu^4\psi_5\bigr)=0\quad,}\eqno(\z)$$

\noindent{(once} (3.16) has been used), from which one gets

$$\psi_5={\la 13\ra\la 23\ra\over\mu^4}\psi_3+{\la
13\ra\over\mu^2}\psi_2+\psi_1\quad.\eqno(\z)$$

The next constraint, after using (3.18), gives

$$\dot\Omega_4=\bigl\{\Omega_4,{\cal H}_T\bigr\}_{PB}\equiv\Omega_5=\la
13\ra\la 12\ra\pi_1-\mu^2\la 13\ra\pi_2-\mu^4\pi_3+2{\mu^6\over
2M}\mu^4\zeta=0\quad,\eqno(\z)$$

\noindent{and} $\zeta$ can be obtained as a function of $\pi_1\,$,
$ \pi_2\,$, and $\pi_3\,$, thus bringing the generation of
secondary constraints to and end.

${\cal H}_T$ being quadratic in $\psi$'s and $\pi$'s, guarantees
an alternance of linear constraints involving the fields and the
momenta. In spite of the occurrence of space derivatives in ${\cal
M}$, they cancel out and the constraints are algebraic. From this
set of constraints, the multipliers $\psi_4$ and $\psi_5\,$,
together with their conjugate momenta $\pi_4$ and $\pi_5\,$, can
be worked out.

The Lagrangian (3.5) can be expressed in terms of the independent
variables $\psi_\alpha\quad (\alpha=1,2,3)\,$. Notice that
implementing these algebraic constraints in ${\cal L}^6$ does not
modify the second differential order already attained in (3.5).
One obtains

$${\cal L}^6=\dot\psi_{\alpha}{\bar{\cal
K}}_{\alpha\beta}\dot\psi_{\beta}+\psi_{\alpha}{\bar{\cal
M}}_{\alpha\beta}\psi_{\beta}-j\psi_3\quad,\eqno(\z)$$

\noindent{where}

$${\bar{\cal K}}_{\alpha\beta}\equiv{1\over 2}\bigl({\cal
K}_{\alpha\beta}+{\cal K}_{\alpha B}{\cal N}_{B\beta}+{\cal
N}_{\alpha A}{\cal K}_{A\beta}\bigr)={\mu^6\over
2M}\left(\matrix{0&0&1\cr0&1&{\la 13\ra\over\mu^2}\cr1&{\la
13\ra\over\mu^2}&{\la 13\ra\la
23\ra\over\mu^4}\cr}\right)\quad,\eqno(\z)$$

$$\eqalign{{\bar{\cal M}}_{\alpha\beta}&\equiv{1\over 2}\bigl({\cal
M}_{\alpha\beta}+{\cal M}_{\alpha B}{\cal N}_{B\beta}+{\cal
N}_{\alpha A}{\cal M}_{A\beta}\bigr)=\cr &={\mu^6\over
2M}\left(\matrix{0&\mu^2&- M_3^2\cr\mu^2&\la 13\ra -M_2^2&-{\la
13\ra\over\mu^2}M_3^2\cr-M_3^2&-{\la 13\ra\over\mu^2}M_3^2&-{\la
13\ra\la  23\ra\over\mu^4}M_3^2\cr}\right)\quad,\cr}\eqno(\z)$$

\noindent{with} $\alpha,\beta,\dots=1,2,3\,$; $A,B,\dots=4,5\,$;
and

$${\cal N}_{A\beta}\equiv\left(\matrix{0&1&{\la 13\ra\over\mu^2}\cr1&{\la
13\ra\over\mu^2}&{\la 13\ra\la 23\ra\over\mu^4}\cr}\right)\eqno(\z)$$

\noindent{that} allows to embody (3.16) and (3.20) in the closed form

$$\psi_A = {\cal N}_{A\beta}\psi_{\beta}\quad.\eqno(\z)$$

The symmetric matrices $\bar{\cal K}$ and $\bar{\cal M}$ can be
simultaneously diagonalized by the regular transformation

$$\psi_{\alpha}={\cal R}_{\alpha\beta}\phi_{\beta}\quad,\eqno(\z)$$

\noindent{where}

$${\cal R}_{\alpha\beta}\equiv \left(\matrix{{\la 12\ra\la
13\ra\over\mu^4}&0&0\cr -{\la 13\ra\over\mu^2}&-{\la
23\ra\over\mu^2}&0\cr 1&1&1\cr}\right)\quad.\eqno(\z)$$

\noindent{The} reason for this property will be made clear later on.

The 3rd. row of the non-orthogonal matrix ${\cal R}$ in (3.28),
has been chosen so as to yield the source term in (3.2), in which
the source couples to the sum of the LD effective fields. The
remaining six elements are uniquely determined by requiring ${\cal
R}$ to diagonalize $\bar{\cal K}$ and $\bar{\cal M}$.

The diagonalized matrices are

$${\cal R}^T\bar{\cal K}{\cal R}={\mu^6\over 2M}\, diag\,\biggl({\la 12\ra\la
13\ra\over\mu^4},-{\la 12\ra\la 23\ra\over\mu^4},{\la 13\ra\la
23\ra\over\mu^4}\biggr)\quad,\eqno(\z)$$

$${\cal R}^T\bar{\cal M}{\cal R}={\mu^6\over 2M}\, diag\,\biggl(-M_1^2{\la
12\ra\la 13\ra\over\mu^4},M_2^2{\la 12\ra\la
23\ra\over\mu^4},-M_3^2{\la 13\ra\la
23\ra\over\mu^4}\biggr)\quad,\eqno(\z)$$

\noindent{so} that (3.22) finally writes as ${\cal L}^2$ in (3.2).

This shows again the Ostrogradski-based result [9] of the
equivalence between the six-derivative theory (3.1) and the LD
version (3.2) that reproduces the propagator structure (3.3).

\vfill
\eject


\sectio{\bf Theories with arbitrary n}
\bigskip
The general Lagrangian

$${\cal L}^{2n}=-{1\over 2}{\mu^{d}\over M}\phi\kg 1\gk\kg 2\gk\dots\kg
n\gk\phi -j\phi\quad,\eqno(\z)$$

\noindent{where} $M\equiv \prod\limits_{i<j}\la ij\ra\,$, and
$d=n(n-3)+2$ for dimensional convenience, can be dealt with along
similar lines. The 2-derivative constrained recasting of (4.1) is

$${\cal L}^{2n}={1\over 2}{\mu^{d}\over M}\bigl[-\bar{\psi}_n\kg
1\gk\bar{\psi}_1+\lambda_1(\bar{\psi}_1 -\kg 2\gk\bar{\psi}_2)
+\dots +\lambda_{n-1}(\bar{\psi}_{n-1}-\kg
n\gk\bar{\psi}_n)\bigr]- j\bar{\psi}_n\quad,\eqno(\z)$$

\noindent{with} $\bar{\psi}_n\equiv\phi\,$, and
$\lambda_1,\dots,\lambda_{n-1}$ being Lagrange's multipliers. In
order to have a more compact notation we define

$$\eqalign{\psi_{\alpha} &=\mu^{-2(n-\alpha)}\bar{\psi}_{\alpha}\quad;\quad\alpha=1,\dots ,n\cr\psi_A &=
\mu^{-2\alpha}\lambda_{\alpha}\quad\,\,\,\,\quad;\quad
A=n+\alpha\, ;\, \alpha=1,\dots ,n- 1}\eqno(\z)$$

\noindent{so} that $[\psi_i]=1$\quad$(i=1,\dots ,2n-1)$. Then

$${\cal L}^{2n}={1\over 2}\dot{\Psi}^T{\cal K}\dot{\Psi}+{1\over 2}\Psi^T{\cal
M}\Psi-J^T\Psi\eqno(\z)$$

\noindent{with} $J_i=j\delta_{in}$, and the $(2n-1)\times(2n- 1)$
matrices ${\cal K}$ and ${\cal M}$ are given by

$$\eqalign{{\cal K}_{ij}&\equiv\sigma(\delta_{i,j-n+1}+\delta_{j,i-
n+1})\cr{\cal M}_{ij}&\equiv\sigma\bigl[-(M_{{\underline
i}}^2\delta_{i,j- n+1}+M_{{\underline
j}}^2\delta_{j,i-n+1})+\mu^2(\delta_{i,j-n}+\delta_{j,i-
n})\bigr]\quad,}\eqno(\z)$$

\noindent{with} $\sigma\equiv{\mu^{n(n-1)}\over 2M}$. This``mass"
matrix contains again space derivatives. Here and in the following
an underlined index means that Einstein summation convention does
not apply. The canonical conjugate momenta are now

$$\pi_i={\partial{\cal L}^{2n}\over\partial\dot{\psi}_i}\quad,\eqno(\z)$$

\noindent{i}.e., in closed notation,

$$\Pi={\cal K}\dot\Psi\quad.\eqno(\z)$$

Defining the matrix ${\cal K}^{'}$

$${\cal K}^{'}_{ab}={1\over\sigma}{\cal K}_{ab}\quad\quad (a,b=1,\dots,2n-
2)\quad,\eqno(\z)$$

\noindent{one} sees that det$\,{\cal K}^{'}\ne 0$, while
det$\,{\cal K}=0$. This means that we only have one primary
constraint, namely

$$\Omega_1\equiv\pi_{2n-1}-\pi_1=0\quad.\eqno(\z)$$

\vfill
\eject

\noindent{Then} $\dot{\psi}_{2n-1}$ is not worked out, while
$\dot{\psi}_a\quad(a=1,\dots,2n-2)$ can be expressed in terms of
$\pi_a$ and $\dot{\psi}_{2n-1}$. The first $2n- 2$ components of
eq.(4.7), namely

$$\pi_a=\sigma{\cal K}^{'}_{ab}\dot{\psi}_b+\sigma\delta_{an}\dot{\psi}_{2n-1}\eqno(\z)$$

\noindent{give}

$$\dot{\psi}_a={1\over\sigma}{\cal K}^{'}_{ab}\pi_b-\delta_{a1}\dot{\psi}_{2n-1}\quad.\eqno(\z)$$

After checking that the terms in $\dot{\psi}_{2n-1}$ cancel out,
the Hamiltonian has the simple expression

$${\cal H}={1\over 2}\sigma\pi_a{\cal K}_{ab}^{'}\pi_b-{1\over 2}\psi_i{\cal
M}_{ij}\psi_j+j\psi_n\quad.\eqno(\z)$$

In ${\cal H}$ only $2n-2$ momenta $\pi_a$ occur against $2n-1$
fields $\psi_i\,$, because of the primary constraint (4.9). One
may restore the dependence on $2n-1$ momenta by introducing the
``total Hamiltonian"

$${\cal H}_T={\cal H}+\zeta\Omega_1\quad,\eqno(\z)$$

\noindent{where} $\zeta$ is a Lagrange multiplier.

From the stability condition on $\Omega_1\,$, a cascade of
secondary constraints follows, eventually ending with an equation
that determines the value of $\zeta$. We outline here the steps
closely following the lines of section 3.

$$\dot{\Omega}_1=\bigl\{\Omega_1,{\cal H}_T\bigr\}_{PB}\equiv\Omega_2=0\quad\Rightarrow\quad\psi_{n+1}=\psi_{n-
1}+{\la 1n\ra\over\mu^2}\psi_n\quad.\eqno(\z)$$

\noindent{Then}

$$\dot{\Omega}_2=\bigl\{\Omega_2,{\cal
H}_T\bigr\}_{PB}\equiv\Omega_3=\mu^2\pi_{2n-2}+\la
1n\ra\pi_1-\mu^2\pi_2(1-
\delta_{n2})-2\sigma\zeta\delta_{n2}=0\quad.\eqno(\z)$$

\noindent{If} $n=2$, eq.(4.15) gives $\zeta$ in terms of $\pi_1$
and $\pi_2\,$, and the cascade stops here, but if $n>2$ it yields

$$\pi_{2n-2}=-{\la 1n\ra\over\mu^2}\pi_1+\pi_2\quad.\eqno(\z)$$

The next step is $\dot{\Omega}_3=\bigl\{\Omega_3,{\cal
H}_T\bigr\}_{PB}\equiv\Omega_4=0$, which together with (4.14)
gives

$$\psi_{n+2}=\psi_{n-2}+{1\over\mu^2}(\la 1n\ra+\la2,n-1\ra)\psi_{n-
1}+{1\over\mu^4}\la 1n\ra\la 2n\ra\psi_n\quad,\eqno(\z)$$

\vfill
\eject

\noindent{and}, proceeding further, we obtain for the momenta

$$\eqalign{\dot{\Omega}_4&=\bigl\{\Omega_4,{\cal
H}_T\bigr\}_{PB}\equiv\Omega_5=\mu^4\pi_{2n-3}-\la 1n\ra\la 1,n-
1\ra\pi_1+\cr&+\mu^2(\la 1n\ra+\la
2,n-1\ra)\pi_2-\mu^4\pi_3(1-\delta_{n3})-
2\sigma\mu^4\zeta\delta_{n3}=0\quad,}\eqno(\z)$$

\noindent{where} (4.16) has been taken into account. Again, if
$n=3$, the process stops here and we have reproduced the results
of section 3. If $n>3$, eq.(4.18) yields

$$\pi_{2n-3}={1\over\mu^4}\la 1n\ra\la 1,n-1\ra\pi_1-{1\over\mu^2}(\la
1n\ra\la2,n-1\ra)\pi_2+\pi_3\quad,\eqno(\z)$$

\noindent{and} the process goes on.

For illustrative purposes, we complete here the steps that cover
the case $n=4$. $\dot{\Omega}_5=\bigl\{\Omega_5,{\cal
H}_T\bigr\}_{PB}\equiv\Omega_6=0$, yields

$$\eqalign{\psi_{n+3}&=\psi_{n-3}+{1\over\mu^2}(\la 3,n-2\ra+\la 2,n-
1\ra+\la1n\ra)\psi_{n-2}+\cr&+{1\over\mu^4}\bigl(\la 2,n-1\ra\la
3,n-1\ra+ \la 1n\ra(\la 2,n-1\ra+\la
3n\ra)\bigr)\psi_{n-1}+\cr&+{1\over\mu^6}\la 1n\ra\la 2n\ra\la
3n\ra\psi_n\quad,}\eqno(\z)$$

\noindent{and} $\dot{\Omega}_6=\bigl\{\Omega_6,{\cal
H}_T\bigr\}_{PB}\equiv\Omega_7=0$ gives $\zeta$ in terms of
$\pi_1\,$, $\pi_2\,$, $\pi_3\,$, and $\pi_4\,$.

In general, for a fixed $n$, the quadratic dependence of ${\cal
H}$ on $\pi_i$ and $\psi_i\,$, together with the primary
constraint $\Omega_1\,$, leads to a set of secondary constraints
$\Omega_k$ that splits in two classes according to $k$ being even
or odd. A constraint $\Omega_{2j}\quad(j=1,\dots,n-1)$ is a linear
combination of $\psi_i$ and gives $\psi_{n+j}$ in terms of
$\psi_n,\dots,\psi_{n-j}$. A constraint $\Omega_{2j-
1}\quad(j=2,\dots,n-1)$ is linear in $\pi_i$ and gives
$\pi_{2n-j}$ in terms of $\pi_1,\dots,\pi_j\,$. Finally,
$\Omega_{2n-1}$ fixes the value of $\zeta$ and stops the process.

One can prove that the constraints on the momenta $\Omega_{2n-1}$
do not contain space derivatives, even though the elements of
${\cal M}$ involved in their calculation contain the Laplacian
operator. This will be shown in the Appendix A.

Like in (3.26), we take

$$\psi_A={\cal N}_{A\beta}\psi_\beta\eqno(\z)$$

\noindent{with} indices $\alpha,\beta,\dots=1,\dots,n$ and
$A,B,\dots=n+1,\dots,2n-1$. \vfill \eject

Now ${\cal N}$ is a $(n-1)\times n$ numerical matrix whose three
first rows can be read from (4.14), (4.17) and (4.20). In order to
write the elements of ${\cal N}\,$, it is useful to introduce the
indices $s=A-n , (s=1,...,n-1)\,$, that labels the rows of ${\cal
N}$, and $p= A+{\beta}-2n={\beta}+s-n , (p=1+s-n,...,s)$, that
indicates the odd-diagonals of ${\cal N}$. Then, with
$P^{s+1}_{p+1}\equiv{\cal N}_{n+s,n-s+p}$, one has:

$$\eqalign{P^{s+1}_{p+1}&=0\quad {\rm for}\quad p<0\quad;\quad P^{s+1}_{1}=1\quad;\quad
P^{s+1}_{2}={1\over {\mu^2}} \sum \limits^{s}_{j_{1}=1}\la
j_1,n+1-j_1\ra\cr
P^{s+1}_3&={1\over{\mu^4}}\sum\limits^{s-1}_{j_1=1}\la
j_1,n+1-j_1\ra\sum\limits^{s-j_1}_{j_2=1}\la
j_1+j_2,n+2-j_1-j_2\ra\cr
&.........................................................................................\cr
P^{s+1}_{p+1}&={1\over{\mu^{2p}}}\sum\limits^{s-p+1}_{j_{1}=1}\la
j_{1},n+1-j_{1}\ra\sum\limits^{s-p+2-j_{1}}_{j_{2}=1}\la
j_{1}+j_{2},n+2-j_{1}-j_{2}\ra \dots\cr
&\dots\sum\limits^{s-p+l-j_{1}-...-j_{l-1}}_{j_{l}=1}\la
j_{1}+...+j_{l},n+l-j_{1}-...-j_{l}\ra \dots\cr
&\dots\sum\limits^{s+1-j_{1}-...-j_{p-2}}_{j_{p-1}=1}\la
j_{1}+...+j_{p-1},n+p-1-j_{1}-...-j_{p-1}\ra\cr
\quad\quad\quad&\sum\limits^{s-j_{1}-...-j_{p-1}}_{j_{p}=1}\la
j_{1}+...+j_{p},n+p-j_{1}-...-j_{p}\ra\cr
&.........................................................................................\cr
P^{s+1}_{s}&={1\over{\mu ^{2(s-1)}}}\sum\limits^{2}_{j_{1}=1}\la
j_{1},n+1-j_{1}\ra\sum\limits^{3-j_{1}}_{j_{2}=1}\la
j_{1}+j_{2},n+2-j_{1}-j_{2}\ra\dots\cr
&\dots\sum\limits^{s-1-j_{1}-...-j_{s-3}}_{j_{s-2}=1}\la
j_{1}+...+j_{s-2},n+s-2-j_{1}-...-j_{s-2}\ra\cr
\quad\quad\quad&\sum\limits^{s-j_{1}-...-j_{s-2}}_{j_{s-1}=1}\la
j_{1}+...+j_{s-1},n+s-1-j_{1}-...-j_{s-1}\ra\cr
P^{s+1}_{s+1}&={1\over{\mu^{2s}}}\la 1n\ra\la 2n\ra ...\la
sn\ra}\eqno(\z)$$

\noindent{and} (4.21) can be rewritten as

$$\psi_{n+s}=\sum\limits^{s}_{p=0}P^{s+1}_{p+1}\psi_{n-s+p}\eqno(\z)$$

The proof of (4.22) and (4.23) is given in Appendix B, were the relation

$$P^{s+1}_{p+1}=P^{n+p-s}_{p+1}\quad\quad {\rm for}\quad p>0\eqno(\z)$$

\noindent{is} also proven. Taking $P_{\alpha\beta}=P^{s+1}_{p+1}$,
with $\alpha =s+1\, ,\, \beta=(p+1)-(s+1)+n=p-s+n$, and
$P_{1n}=1\,,$  $P_{1\beta}=0$ for $\beta<n$, we can consider
$P_{\alpha\beta}$ as a $n\times n$ matrix, that is nothing more
than the matrix $\cal N$ enlarged with a first row, which is
symmetric as a consequence of (4.24).

After using the constraints (4.23) to keep only the independent variables,
the Lagrangian again is

$${\cal L}^{2n}=\dot{\psi}_\alpha\bar{\cal
K}_{\alpha\beta}\dot{\psi}_\beta+\psi_\alpha\bar{\cal
M}_{\alpha\beta}\psi_\beta-j\psi_n\quad.\eqno(\z)$$

The $n\times n$ matrices ${\bar{\cal K}}$ and ${\bar{\cal M}}$
have the same structure in terms of ${\cal K}\,$, ${\cal M}$ and
${\cal N}$ given in (3.23) and (3.24). Taking into account (4.22),
one checks that

$$\bar{\cal K}_{\alpha\beta}={\mu^{n(n-1)}\over 2M}P^{\alpha}_{\beta -n+\alpha}=
{\mu^{n(n-1)}\over 2M}P_{\alpha\beta}=\bar{\cal K}_{\beta\alpha}\eqno(\z)$$

\noindent{is} the generalization of (3.23). For $\bar{\cal M}$ one
has two contributions,
 with and without the operators $M^2_i\,$. The first one can be written, for $\alpha\le\beta$ as

$$\bar{\cal M}_{\alpha\beta}=-{\mu^{n(n-1)}\over 2M}M^{2}_{\underline \beta  }P_{\alpha\beta\quad,}\eqno(\z)$$

\noindent{which} displays $M^2_{\underline\alpha}$ for
$\beta\le\alpha$.

The contribution without $M^2_i$ is given by

$$\eqalign{\bar{\cal M}_{\alpha\beta}&=0
\quad\quad\quad\quad\quad\quad\quad\quad {\rm if}\quad\alpha+\beta<n\cr
\bar{\cal M}_{\alpha\beta}&={\mu^{n(n-1)}\over 2M}\mu^2P_{\alpha
,\beta +1}\quad {\rm if}\quad\alpha+\beta\ge
n\,\,;\,\,\alpha,\beta\not= n\cr \bar{\cal
M}_{\alpha\beta}&=0\quad\quad\quad\quad\quad\quad\quad\quad {\rm
if}\quad\alpha\,\,{\rm or}\,\,\beta =n\quad .}\eqno(\z)$$

\noindent{The} formulas (4.27) and (4.28), are the generalization
of (3.24) for arbitrary $n\,$.

The diagonalization of (4.25) will be accomplished, as in (3.27),
by a $n\times n$ real matrix $\cal R$. We again impose ${\cal
R}_{n\beta}=1\,, (\beta=1,\dots,n)$ to ensure that the current
couples to each one of the degrees of freedom. The requirement of
simultaneously diagonalizing ${\bar{\cal K}}$ and ${\bar{\cal
M}}$, yields $n(n-1)$ quadratic equations that determine the $n(n-
1)$ remaining elements of ${\cal R}$. The existence of such a
regular $\cal R$ with real elements is guaranteed by the
underlying Lorentz covariance. In fact, the constraints used to
get from (4.4) to (4.25) are the (covariant) ones on the fields.
In the next section we will see that they can be obtained also
from the (covariant) field equations of (4.2). They can be
implemented on this Lagrangian which can be then diagonalized
directly avoiding the decomposition 3+1. The diagonalizing matrix
turns out to be

$$\eqalign{{\cal R}_{\alpha\beta}&=1\quad\quad;\quad\quad (\alpha=n)\cr
      {\cal R}_{\alpha\beta}&=(-1)^{n-\alpha}\mu^{-2(n-
\alpha)}\la\beta,\alpha+1\ra\la\beta,\alpha+2\ra\dots\la\beta,n\ra\quad;
\quad (\beta\le\alpha <n)\cr
      {\cal R}_{\alpha\beta}&=0\quad\quad;\quad\quad (\alpha <\beta) \quad,}\eqno(\z)$$

\noindent{as} is shown in the Appendix C. Of course, for $n=3$,
this $\cal R$ is just (3.28).

The diagonalized matrices are now the generalization of (3.29) and (3.30)

$$\eqalign{{\cal R}^T\bar{\cal K}{\cal R}&={\mu^{n(n-1)}\over 2M}\,
diag\,\biggl((-1)^{n-1}{\la 12\ra\la
13\ra ...\la 1n\ra\over\mu^{2(n-1)}},(-1)^{n-2}{\la 12\ra\la 23\ra
...\la 2n\ra\over\mu^{2(n-1)}},...\cr ...&,(-1)^{n-i}{\la 12\ra\la
13\ra ...\la 1,i\ra\la i,i+1\ra ...\la
i,n\ra\over\mu^{2(n-1)}},... ,{\la 1n\ra\la 2n\ra ...\la
n-1,n\ra\over\mu^{2(n-1)}}\biggr)\,\, ,}\eqno(\z)$$

$$\eqalign{{\cal R}^T\bar{\cal M}{\cal R}&={\mu^{n(n-1)}\over 2M}\, diag\,\biggl((-1)^{n}M_1^2{\la
12\ra...\la 1n\ra\over\mu^{2(n-1)}},(-1)^{n-1}M_2^2{\la 12\ra\la
23\ra...\la 2n\ra\over\mu^{2(n-1)}},\cr ...,&(-1)^{n-i+1}M_i^2{\la
12\ra...\la 1i\ra\la i,i+1\ra...\la
in\ra\over\mu^{2(n-1)}},...,-M^2_n{\la 1n\ra...\la
n-1,n\ra\over\mu^{2(n-1)}}\biggr)\, .}\eqno(\z)$$

With $\psi_{\alpha}=\cal R_{\alpha\beta}\phi_{\beta}\,$, and
integrating by parts, the final Lagrangian reads

$$\eqalign{{\cal L}^{2n}&=(-1)^n{1\over 2}{1\over\la 1\ra}\phi_1\kg 1\gk\phi_1+(-1)^{n-1}{1\over
2}{1\over\la 2\ra}\phi_2\kg 2\gk\phi_2+\dots \cr
&\dots+(-1)^{n-i+1}{1\over 2}{1\over\la i\ra}\phi_i\kg
i\gk\phi_i+\dots-{1\over 2}{1\over\la n\ra}\phi_n\kg n\gk\phi_n-
j(\phi_1+\dots+\phi_n)\quad,}\eqno(\z)$$

\noindent{where} $\la i\ra\equiv{1\over\mu^{n(n-1)}}M
\prod\limits_{j\ne i}{1\over{\vert\la ij\ra\vert}}$. This is the
result expected from the covariant Ostrogradski method shown in
[9].

\vfill
\eject


\sectio{\bf The covariant phase space}
\bigskip

The constraints on the fields and momenta obtained by the Dirac's
method define a subset of the phase space in which the solutions
to the equations of motion are contained. In non pathological
situations, there is a one to one correspondence between points of
this subspace and these solutions [12]. In the system under study
we have found algebraic constraints on the fields that can be
derived also directly from the equations of motion. They define
subsets of the configuration space in which the trajectories of
the system lie. Now we easily see how these constraints are
derived (we put $j=0$ for simplicity).

From (4.4) the following equations of motion are obtained respectively for the fields
 $\psi_{n+i}\,$, $\psi_i\,$ $(1\leq i < n)\,$ and $\psi_n\,$.

$$ \eqalign{ \mu^2\psi_i &= \kg i+1\gk \psi_{i+1}  \cr
              \mu^2\psi_{n+i}&=\kg i\gk \psi_{n+i-1}  \cr
              0&= \kg 1\gk \psi_1 + \kg n\gk \psi_{2n-1} } \eqno(\z) $$

From them it is possible to work out the $n-1$ fields $\psi_i \; (n < i <2n)$ in terms of
the $n$ independent fields $\psi_i \; (1 \leq i \leq n)\,$. This is done by using the
first of the equations (5.1) to obtain

$$ \square\,\psi_{i+1}= \mu^2\psi_i + m^2_{i+1}\psi_{i+1}\quad (1\leq i<n)\eqno(\z)$$

\noindent{Then} iterative use can be made of the second of the
equations (5.1), taking (5.2) into account in each step, to obtain
the algebraic constraints on the fields (4.14), (4.17) and their
general form (4.23). The action (4.25), already depending only on
the independent fields, explicitly reads:

$${\cal L}^{2n} = {1\over 2}[-\psi_n \kg 1\gk \psi_1 + \sum\limits^{n-1}_{s=1}
                \sum\limits^{s}_{p=0} P^{s+1}_{p+1}
            \psi_{n-s+p}(\mu^2 \psi_s - \kg s+1 \gk)\psi_{s+1} ] \eqno(\z)$$

\noindent{where} the non-diagonal structure is apparent.

The diagonalization matrix (4.29) is suggested by the equations of motion stemming
from (5.3) and directly derivable also from (5.1):

$$ \eqalign{ \mu^{2(n-1)}\psi_1 &= \kg 2\gk \cdot\cdot\cdot \kg n\gk \,\psi_n  \cr
              \mu^{2(n-2)}\psi_2 &= \kg 3\gk \cdot\cdot\cdot \kg n\gk \,\psi_n  \cr
              .......\; & \;\quad  .......... \cr
              \mu^2\psi_{n-1}&= \kg n\gk \,\psi_n } \eqno(\z) $$

\noindent{Defining} $n$ new fields $\phi_a \; (1\leq a \leq n)\,$,
with $\psi_n = \sum\limits^n_{a=1}\phi_a $ and obeying $\kg a \gk
\phi_a = 0\;$, the equations of motion (5.4) become

$$ \eqalign{ \mu^{2(n-1)}\psi_1
       &= \langle 2,1 \rangle \cdot\cdot\cdot \langle n,1 \rangle\,\phi_1  \cr
              \mu^{2(n-2)}\psi_2
      &= \langle 3,1 \rangle \cdot\cdot\cdot \langle n,1 \rangle\,\phi_1
         + \langle 3,2 \rangle \cdot\cdot\cdot \langle n,2 \rangle\,\phi_2 \cr
              .......\; & \;\quad  .......... \cr
              \mu^2\psi_{n-1}
        &= \langle n,1 \rangle \,\phi_1 +\cdot\cdot\cdot
                               + \langle n,n-1 \rangle \,\phi_{n-1} \cr
        \psi_n &= \phi_1 + \cdot\cdot\cdot +\phi_n } \eqno(\z) $$

\noindent{From} these equations it is immediate to read out the elements of the
diagonalizing matrix (4.29).

The procedure implemented in this section provides a useful
shortcut to get the diagonalized theory. However it is worth
noticing that one can not avoid the arduous path followed in
section 4 because it guarantees the consistency and the stability
of the constraints. Moreover, the Dirac's procedure (when it
exists) yields the structure of the reduced phase space (null set
of all the constraints, modulo gauge, equipped with the induced
non-degenerate symplectic form). The knowledge of the reduced
phase space is necessary, for example, when one is interested in
using path integral methods because it is on this space where the
canonical quantization takes place. Then the covariant method
presented in this section complements the results obtained in
section 4 besides being specially suitable for the DOF
diagonalization.

\vfill
\eject


\sectio{\bf Applications to other theories}

\bigskip

The constraint method we have developed for scalar theories can be implemented for HD
vector and tensor theories as well. However the canonical 3+1 procedure described
in Section 4 is not convenient for practical applications,
mainly when more complex theories as these are considered.
The simplified covariant version introduced in Section 5
gets rid of the intermediate constraints
$\Omega_{2j+1}$ on the (non Lorentz-covariant) momenta and obtains
the useful ones $\Omega_{2j}$ on the (covariant) fields by
deriving them directly from the (always covariant) equations of
motion for all the fields (original HD, auxiliary and multipliers).
Moreover when gauge symmetries occur there are further constraints,
namely the first class ones, that make more cumbersome
the process of working out the Lagrange multipliers.
In any case, the covariant phase-space is the method of choice.

\bigskip

\centerline{\bf HD vectors and differential-forms}

In the case of HD vector theories one can incorporate gauge
symmetries to the discussion.  A typical example is the
generalized QED proposed by F.Bopp and B.Podolsky [13]

$$\eqalign{{\cal L} = - {1\over 4}F_{\mu\nu}F^{\mu\nu} &- {1\over 4m^2}F_{\mu\nu}\,\square\,
F^{\mu\nu} - j_{\mu}A^{\mu} \;.}\eqno(\z)$$

\noindent{The} structure of its constraints, in the gauge-fixed
case, has been studied in [3] by a
 canonical forcefully non-covariant analysis carried out on the Ostrogradski-based
 order-reduction procedure.

A recasting of a higher-derivative gauge-invariant Yang-Mills
theory as a two-derivative one by means of constraints has been
done in a non-covariant 3+1 way [14]. Of course, in the non gauge-fixed theory,
second class constraints arise which coexist harmlessly
with the first class relevant ones.  However we are interested in
keeping the explicit Lorentz covariance and, to this
end, we make use of the same algebraic manipulations introduced in
section 5.

The theory described by (6.1) can be considered a member of a wider family of
theories which share the same general structure of the scalar models (4.1)

$${\cal L}^{2n}=-{1\over 2}{\mu^{d}\over M}\phi\wedge * \kg 1\gk\kg 2\gk\dots\kg
n\gk\phi -j\wedge*\phi\quad,\eqno(\z)$$

\noindent{where} $\phi$ is a $s$-form (1-form in the generalized
electromagnetism), $\wedge$ is the exterior product, $*$ the Hodge
dual (associated to the Minkowski metric), and now the $\kg i\gk$
operators must be understood as combinations of the exterior
derivative $d$ and its dual $\delta$ such as

$$
\kg i\gk \equiv \delta d+ m_i^2\,.
$$

Notice that if $m_n=0$ the Lagrangian (6.2) describes a gauge
theory with gauge transformations
$\phi\mapsto \phi+d\Lambda\,$, where $\Lambda$ is an arbitrary $(s-1)$-form.

With all these considerations, the field equations derived from
(6.2) have the same look of equations (5.1) for the scalar model.
Moreover, all the arguments that follow this equations in section
5 apply in the same way to the (6.2) HD theory that depends now on
differential-form fields. Notice that, despite the gauge
invariance of the theory, the order reduction method and the
linear redefinition that diagonalize the DOF in the Lagrangian
work  exactly as in the scalar case.

\bigskip

\centerline{\bf HD Gravity}

The covariant Ostrogradski order-reduction of the four-derivative
gravity leads to a two-derivative equivalent in which the particle
DOF can be fully diagonalized in both the Diff-invariant case [8]
and with a gauge fixing [15].

The constraint technique for the order-reduction of a pure
four-derivative conformally invariant gravitational Lagrangian has
been already used in a 3+1 non-covariant form [16], where further
first class constraints from Diff-invariance occur. In a covariant
treatment and for the general case including also two-derivative
terms [17], a seemingly similar method is adopted where in place
of the Lagrange multiplier a less trivial auxiliary field
featuring a squared (mass)term is used. A little work shows
however that this method is identical to the covariant
Ostrogradski's [9]. We illustrate this on the grounds of the
scalar model in the Appendix D.

The covariant constraint method introduced in this paper provides
a new approach. The most immediate application in
higher-derivative gravity regards the linearized theory, usually
considered when analyzing the DOF. Take for example the
four-derivative Lagrangian

$$ {\cal L} = \sqrt{-g} [ aR+bR^2+cR_{\mu\nu}R^{\mu\nu} ]\;. \eqno(\z) $$

\noindent{The} linearization around the flat Minkowski metric, namely
$\;g_{\mu\nu}=\eta_{\mu\nu}+h_{\mu\nu}\;$, simplifies it to

$$
{\cal L}= {{a}\over{2}}h_{\mu\nu}G^{\mu\nu\,
\alpha\beta}r_{\alpha\beta}+r_{\mu\nu}Q^{\mu\nu\,\alpha\beta}(b,c)
r_{\alpha\beta}\;,  \eqno(\z)$$

\noindent{where} $r_{\mu\nu}$  comes from the linearization of the
Ricci tensor

$$
r_{\mu\nu}={\cal R}_{\mu\nu\, \alpha\beta}h^{\alpha\beta}\equiv
{1\over
2}\left[h^{\alpha}\,_{\mu,\nu\alpha}+h^{\alpha}\,_{\nu,\mu\alpha}
-\square h_{\mu\nu}-h^{\alpha}\,_{\alpha,\mu\nu} \right]\,,
$$

\noindent{and} $G^{\mu\nu\, \alpha\beta}$ and $ Q^{\mu\nu\,
\alpha\beta}$ are numerical matrices:

$$\eqalign{
 G^{\mu\nu\, \alpha\beta}&\equiv {1\over 2}\left[
\eta^{\mu\nu}\eta^{\alpha\beta}-\eta^{\mu\alpha}\eta^{\nu\beta}
-\eta^{\mu\beta}\eta^{\nu\alpha} \right]\quad ,\cr  {\cal
Q}^{\mu\nu\,\alpha\beta}(b,c)&\equiv
b\,\eta^{\mu\nu}\eta^{\alpha\beta} +{c\over
2}\left[\eta^{\mu\alpha}\eta^{\nu\beta}+\eta^{\mu\beta}\eta^{\nu\alpha}\right]\,.}
$$

\noindent{The} field equation for $h_{\mu\nu}$ can be written
(omitting indices) in terms of the objects above in the form

$$
\left[G{\cal R}+{a\over 2}GQ^{-1}G\right]G{\cal R} h=0\,,\eqno(\z)
$$

\noindent{Its} straightforward to show that the general solution
to (6.5) is

$$
h={\rm h}_1+{\rm h}_2\,,\eqno(\z)
$$

\noindent{where} ${\rm h}_1$ and ${\rm h}_2$ satisfy, respectively

$$
G{\cal R}\,{\rm h}_1=0\quad;\quad G{\cal R}\, {\rm h}_2=-{a \over
2} GQ^{-1}G\, {\rm h}_2\,.  \eqno(\z)
$$

Omitting indices, the order-reduction of the theory by means of a
Lagrange multiplier yields the two-derivative local Lagrangian

$$
{\cal L}=h_2^{\rm t}\left[{\cal R}^{\rm t}QG+{a\over 2}\right]h_1
+h^{\rm t}_3[\mu^2h_1-G{\cal R} h_2]\,.\eqno(\z)
$$

\noindent{where} $\;h_{1 \mu\nu}\;$ is a new field and $\;h_{3
\mu\nu}\;$ is the multiplier. Of course, because of the
Diff-invariance, first class constraints will remain when the
Dirac procedure is carried out. However, it is possible to avoid
the use of Dirac method if one looks directly at the space of
solutions of the field equations. The field equations for the
Lagrangian (6.8) are

$$\eqalign{
&\left[{\cal R}^{\rm t}QG+{a\over 2}\right]h_1-G{\cal
R}h_3=0\,,\cr &\mu^2h_3+\left[{a\over 2}+GQ{\cal
R}\right]h_2=0\,,\cr &\mu^2h_1-G{\cal R}h_2=0}\eqno(\z)
$$

\noindent{The} last equation leads to  $G{\cal R}h_2=\mu^2h_1$.
Then, the second one expresses algebraically  $h_3$ in terms of
$h_1$ and $h_2$:

$$
h_3=-{a\over{2\mu^2}}h_2-GQGh_1\,.
$$

\noindent{Using} this constraint in the Lagrangian (6.8)  we
obtain

$$
{\cal L}={a\over{2\mu^2}}h_2^{\rm t}G{\cal R}h_2+2h_1^{\rm
t}GQ{\cal R}h_2-\mu^2h_1^{\rm t}GQGh_1\,. \eqno(\z)
$$

\noindent{The} last step is again a diagonalization. The hint is
given by (6.6)-(6.7), identifying $h=h_2$. This leads us to
introduce new fields ${\rm h}_1$, ${\rm h}_2$
$$\eqalign{
h_1 &=-{a\over{2\mu^2}}GQ^{-1}G\,{\rm h}_2\cr h_2 &={\rm h}_1+{\rm
h}_2\,,}
$$

\noindent{in} terms of which the Lagrangian is

$$
{\cal L}^2={a\over{2\mu^2}}\,{\rm h}_1\,G{\cal R}\,{\rm
h}_1-{a\over{2\mu^2}}\,{\rm h}_2\,G{\cal R}\,{\rm h}_2
-{{a^2}\over{4\mu^2}}\,{\rm h}_2\,GQ^{-1}G\,{\rm h}_2\,,
$$

\noindent{where} now is clear that ${\rm h}_1$ propagates the
massless graviton DOF whereas ${\rm h}_2$ carries  a spin 2 with
square mass $-a/c$ and a scalar with square mass  $a/2(c+3b)$.


\bigskip
\bigskip

\sectio{\bf Conclusions}
\bigskip

We have shown how to deal with $2n$-derivative relativistic scalar theories
by writing them directly as second-order constrained Lagrangians with more fields
and suitable Lagrange multipliers. The corresponding canonical conjugate momenta
 are subject to primary constraints, whose conservation in time gives rise to
a finite chain of secondary constraints according to the Dirac's procedure.
Though expected, a non trivial result is that these constraints, later used to
extract the final DOF, are purely algebraic relations that do not involve the
 space derivatives.

Once the constraints have been implemented, we are left with a
second-order Lagrangian for the DOF of the system. We have
performed explicitly the diagonalization for $n=3$, reproducing the result
obtained in [9]. Then the procedure has been generalized to arbitrary $n$, namely (4.29).
This explicit possibility constitutes a definite advantage over Ostrogradski's
method.

The applications to more interesting theories like HD generalized electrodynamics
 and HD Diff-invariant gravity illustrate also the fact that
the order-reducing methods used in the literature fall in two categories:
 the one based in the covariant Ostrogradski and the one based in the constraints
by Lagrange multipliers. The methods based on auxiliary fields with a quadratic
 term, which may look like a variant of the multipliers, actually belong to the
first category and have no obvious manageable extension beyond
the four-derivative order.

In vector and tensor field theories where gauge symmetries occur,
the corresponding first class constraints live together with the second
class ones worked out in this paper and survive the order-reducing procedure
as long as gauge fixings are not considered. The method may then prove useful
 for a detailed analysis of the constraints from gauge (or Diff-)invariance
in these HD theories, chiefly of the fate of the scalar and vector constraints
 of Hamiltonian gravity.
\bigskip
\bigskip

\noindent{\bf Acknowledgements}

Is a pleasure thanks Dr. F.Barbero
for many discussions and comments.

\vfill
\eject


\bigskip

\noindent{\bf Appendix A}
\bigskip

We prove, by induction, that the constraints $\Omega_{2j}\quad
(j=1,\dots,n-1)$ involving the fields, do not contain space
derivatives because the Laplacian operators cancel out.

One first sees, by inspection, that this statement is true for
$\Omega_2\,$: \noindent{the} calculation leading to (4.14) is

$$\Omega_2\equiv\sigma\bigl[\mu^2\psi_{n-1}+(M_1^2-M_n^2)\psi_n-
\mu^2\psi_{n+1}\bigr]=0\quad,$$ \hfill (A.1)

\noindent{where} the cancellation of the Laplacian operator is
apparent, i.e.

$$M_1^2-M_n^2=m_1^2-m_n^2\equiv\la 1n\ra\quad,$$
\hfill (A.2)

\noindent{and} obviously no summation is understood for repeated indices.
Then, let us suppose that, after taking into account the preceding 
constraints, one has that in the constraint

$$\Omega_{2\alpha}=\sigma\bigl[\mu^{2\alpha} \psi_{n-\alpha}+a_1\psi_{n-
\alpha+1}+a_2\psi_{n-\alpha+2}+\dots+a_{\alpha-1}\psi_{n-1}+a_\alpha\psi_n-
\mu^{2\alpha}\psi_{n+\alpha}\bigr]=0\quad,$$
\hfill (A.3)

\noindent{for} $\alpha=1,\dots,n-2$, the 
coefficients $a_1,\dots,a_\alpha$ are real numbers, as are those found 
in (4.14), (4.17) and (4.20). We now prove that this is also true 
for $\Omega_{2\alpha +1}$. In fact

$$\eqalign{\Omega_{2\alpha+1}=\mu^{2\alpha}\pi_{2n-\alpha-1}+a_1\pi_{2n-
\alpha}&+a_2\pi_{2n-\alpha+1}+\dots+a_{\alpha-1}\pi_{2n-2}+a_\alpha\pi_1-
\cr&-\mu^2\pi_{\alpha+1}(1-\delta_{n,\alpha+1})-
2\sigma\mu^2\zeta\delta_{n,\alpha+1}\quad,}$$
\hfill (A.4)

\noindent{from} which

$$\eqalign{\Omega_{2(\alpha+1)}&\equiv\dot\Omega_{2\alpha+1}=\bigl\{\Omega_{2
\alpha+1},{\cal H}_T\bigr\}_{PB}=\cr&=\sigma\bigl[-\mu^{2\alpha}M_{n-
\alpha}^2\psi_{n-\alpha}-a_1M_{n-\alpha+1}^2\psi_{n-\alpha+1}-\dots-
a_{\alpha-1}M_{n-1}^2\psi_{n-1}-\cr&-a_\alpha 
M_n^2\psi_n+\mu^{2\alpha}M_{n+\alpha}^2\psi_{n+\alpha}+\mu^{2(\alpha+1)}\psi_
{n-\alpha-1}+\mu^2a_1\psi_{n-\alpha}+\dots+\cr&+\mu^2a_{\alpha-1}\psi_{n-
2}+\mu^2a_\alpha\psi_{n+1}-
\mu^{2(\alpha+1)}\psi_{n+\alpha+1}\bigl]=0\quad.}$$
\hfill (A.5)
\vfill
\eject

The crucial point now is that, when working $\psi_{n+\alpha}$ out of 
(A.3) and substituting it in (A.5), only differences of squared 
masses $M_i^2$ occur as in (A.2), thus cancelling out the 
operators $\Delta$. Then, by substituting also $\psi_{n+1}$ 
from (A.1), one gets $\psi_{n+\alpha+1}$ as a sum of linear terms 
in $\psi_n,\psi_{n-1},\dots,\psi_{n-\alpha-1}$, the coefficient for 
the last one being the unity. This ends the inductive proof.
\bigskip


\noindent{\bf Appendix B}
\bigskip

We proof (4.23) by induction. First note that for $s=1$ it is
nothing more than (4.14). Assume then that it holds for
$\psi_{n+s}$ with $s\le n-2$. From the Lagrangian (4.4), one has
that the equation of motion for $\psi_{s+1}\,$, $s\le n-2$, is

$$\mu^2\psi_{n+s+1}=\kg s+1\gk\psi_{n+s}=(\square+m^2_{s+1})\psi_{n+s}$$
\hfill (B.1)

\noindent{that}, with (4.23), gives

$$\mu^2\psi_{n+s+1}=\sum\limits^s_{p=0}P^{s+1}_{p+1}\square\,
\psi_{n-s+p}+m^2_{s+1}\sum\limits^s_{p=0}\psi_{n-s+p}\quad,$$
\hfill (B.2)

\noindent{but} the equation of motion for $\psi_{2n-s+p-1}$ is

$$\square\,\psi_{n-s+p}=\mu^2\psi_{n-s+p-1}-m^2_{n-s+p}\psi_{n-s+p}\quad,$$
\hfill (B.3)

\noindent{so}

$$\eqalign{\psi_{n+s+1}&=\sum\limits^{s}_{p=0}{{\la s+1,n-s+p\ra}\over{\mu^2}}
P^{s+1}_{p+1}\psi_{n-s+p}+\sum\limits^{s}_{p=0}P^{s+1}_{p+1}\psi_{n-s+p-1}\cr
&={{\la s+1,n\ra}\over{\mu^2}}P^{s+1}_{s+1}\psi_{n}+\sum\limits^{s-1}_{p=0}
\biggl [{{\la s+1,n-s+p\ra}\over{\mu^2}}P^{s+1}_{p+1}+P^{s+1}_{p+2}\biggr ]\psi_{n-s+p}+\cr
&+P^{s+1}_{1}\psi_{n-s-1}\cr
&={{\la s+1,n\ra}\over{\mu^2}}P^{s+1}_{s+1}\psi_{n}+\sum\limits^{s}_{p=1}
\biggl [{{\la s+1,n-s+p-1\ra}\over{\mu^2}}P^{s+1}_{p}+P^{s+1}_{p+1}\biggr ]\psi_{n-s+p-1}+\cr
&+P^{s+1}_{1}\psi_{n-s-1}\quad ,}$$
\hfill (B.4)

\noindent{and} one gets the desired equation

$$\psi_{n+s+1}=\sum\limits^{s+1}_{p=0}P^{s+2}_{s+1}\psi_{n-(s+1)+p}$$
\hfill (B.5)

\noindent{because} from (4.22) we have

$$\eqalign{&P^{s+1}_{1}=P^{s+2}_1=1\quad\quad;\quad\quad{{\la s+1,n\ra}\over{\mu^2}}
P^{s+1}_{s+1}=P^{s+2}_{s+2}\cr
&{{\la s+1,n-(s+1)+p\ra}\over{\mu^2}}P^{s+1}_p+P^{s+1}_{p+1}=P^{s+2}_{p+1}\quad .}$$
\hfill (B.6)

The third of this relations is checked by splitting the last term
of the sum at the end of $P^{s+2}_{p+1}\,$. With
$j_{p}=s+1-j_{1}-\dots -j_{p-1}\,$, it gives the factor ${\la
s+1,n-(s+1)+p\ra}\over{\mu^2}$ multiplying an expression that can
be rewritten as $P^{s+1}_p\,$. The remaining terms in the sum are
just $P^{s+1}_{p+1}\,$.

Now follows the proof of (4.24). Suppose that $\alpha
=s+1>n+p-s=r+1=\beta$ and write

$$\eqalign{P^{s+1}_{p+1}&={1\over{\mu^{2p}}}\biggl (\sum\limits^{r-p+1}_{j_1=1}
\la j_1,n+1-j_1\ra+\sum\limits^{s-p+1}_{j_1=r-p+2}\la j_1,n+1-j_1\ra\biggr )\times\dots\cr
&\dots\times\biggl (\sum\limits^{r-1-j_1-...-j_{p-2}}_{j_{p-1}=1}\la j_1+...+j_{p-1},n+p-1-j_1-...-j_{p-1}\ra+\cr
&+\sum\limits^{s-1-j_1-...-j_{p-2}}_{j_{p-1}=r-j_1-...-j_{p-2}}\la j_1+...+j_{p-1},n+p-1-j_1-...-j_p\ra\biggr )\times\cr
&\times\biggr (\sum\limits^{r-j_1-...-j_{p-1}}_{j_p=1}\la j_1+...+j_p,n+p-j_1-...-j_p\ra +\cr
&+\sum\limits^{s-j_1-...-j_{p-2}}_{j_p=r+1-j_1-...-j_{p-1}}\la j_1+...+j_p,n+p-j_1-...-j_p\ra\biggr )\quad .}$$
\hfill (B.7)
\vfill
\eject
The second term in the last factor is

$$\la r+1,s\ra +\la r+2,s-1\ra +\dots +\la s-1,r+2\ra +\la s,r+1\ra =0\, .$$
\hfill (B.8)

The biggest superior limit for the first term in this last factor,
when multiplied by the second term in the previous factor, is
determined by the value of $j_{p-1}=r+1-j_1-...-j_{p-2}\,$, so we
have that $j_p=0$ and this product does not appear because
$j_i=0\quad\forall i$. Following this argument, the same happens
with all of the second terms, and only remains the product of the
first ones that is nothing more than $P^{r+1}_{p+1}\,$.


\bigskip
\noindent{\bf Appendix C}
\bigskip

First we will proof (4.30). Taking into account (4.26) and (4.29), we have

$$\bigl ({\cal R}^T\bar{\cal K}{\cal R}\bigr )_{\alpha\beta}=
{{\mu^{n(n-1)}}\over M}\sum\limits^{}_{\rho\ge\alpha ,\sigma\ge\beta}
{\cal R}_{\rho\alpha}P^{\rho}_{\sigma -n+\rho}{\cal R}_{\sigma\beta}\quad.$$
\hfill (C.1)

\noindent{In} what follows we omit the overall factor
${\mu^{n(n-1)}}\over M\,$. Because of (4.22), for $\alpha =\beta
=n$ we get

$$P^n_n={{\la 1n\ra \dots\la n-1,n\ra}\over{\mu^{2(n-1)}}}$$
\hfill (C.2)

\noindent{as} required.

For $\alpha<\beta =n$, we have to prove that
$$\eqalign{&P^n_n +(-1){{\la\alpha ,n\ra}\over{\mu^2}}P^{n-1}_{n-1}+\dots
+(-1)^{n-(\alpha +2)}{{\la\alpha ,\alpha +3\ra...\la\alpha
,n\ra}\over{\mu^{2(n-(\alpha +2))}}}P^{\alpha +2}_{\alpha +2}+\cr
&+(-1)^{n-(\alpha +1)}{{\la\alpha ,\alpha +2\ra...\la\alpha
,n\ra}\over{\mu^{2(n-(\alpha +1))}}}P^{\alpha +1}_{\alpha
+1}+(-1)^{n-\alpha}{{\la\alpha ,\alpha +1\ra...\la\alpha
,n-1\ra\la\alpha ,n\ra}\over{\mu^{2(n-\alpha
)}}}P^{\alpha}_{\alpha}=0.}$$ \hfill (C.3)

In fact ${{\la\alpha ,n\ra}\over{\mu^2}}P^\alpha_\alpha =P^{\alpha
+1}_{\alpha +1}$ because of (B.6), and the two last terms add up
to
$$(-1)^{n-(\alpha +1)}{{\la\alpha,\alpha +2\ra...\la\alpha ,n-1\ra}\over{\mu^{2(n-(\alpha
+2))}}}{{\la\alpha +1,n\ra}\over{\mu^2}}P^{\alpha +1}_{\alpha
+1}\quad,$$\hfill (C.4)

\noindent{that} can be expressed in terms of $P^{\alpha
+2}_{\alpha +2}\,$. Following this procedure step by step we
arrive to

$$P^n_n-{{\la n-1,n\ra}\over{\mu^2}}P^{n-1}_{n-1}=0$$
\hfill (C.5)

\noindent{by} (B.6).

Thus, the n-th column of the kinetic matrix is null but for the element in the n-th row.
Because of the symmetry that gives (4.24), we have for the n-th row

$$\eqalign{&P^n_n+(-1){{\la\alpha ,n\ra}\over{\mu^2}}P^n_{n-1}+\dots
+(-1)^{n-(\alpha +2)}{{\la\alpha ,\alpha +3\ra...\la\alpha ,n\ra}\over{\mu^{2(n-(\alpha +2))}}}P^n_{\alpha +2}+\cr
&+(-1)^{n-(\alpha +1)}{{\la\alpha ,\alpha +2\ra...\la\alpha ,n\ra}\over{\mu^{2(n-(\alpha +1))}}}P^n_{\alpha +1}
+(-1)^{n-\alpha}{{\la\alpha ,\alpha +1\ra...\la\alpha ,n\ra}\over{\mu^{2(n-\alpha)}}}P^n_{\alpha}=0}\quad ,$$
\hfill (C.6)

\noindent{where} $\alpha <n$ indicates now the column.

Take now $\beta\le\alpha <n$. Consider in (C.1) the terms with
$\rho =n$, which are (C.6) with $\beta$ in the place of $\alpha$
and give a null contribution. For $\alpha\le\rho\le n$ we can
write the terms for a given $\rho$ as

$$\eqalign{&{\cal R}_{\rho\alpha}\biggl [P^{\rho}_{\rho}+P^\rho_{\rho -1}(-1){{\la\beta ,n\ra}\over{\mu^2}}
+P^\rho_{\rho -2}(-1)^2{{\la\beta ,n-1\ra\la\beta ,n\ra}\over{\mu^2}}+\dots\cr
&\dots +P^\rho_{\rho -n+(\beta +1)}(-1)^{n-(\beta +1)}{{\la\beta ,\beta +2\ra...\la\beta ,n\ra}\over{\mu^{2(n-(\beta +1))}}}
+\cr
&+P^\rho_{\rho -n+\beta}(-1)^{n-\beta}{{\la\beta ,\beta +1\ra...\la\beta ,n\ra}\over{\mu^{2(n- \beta)}}}\biggr ]\quad.}$$
\hfill (C.7)

\noindent{Applying} (B.6), (C.7) becomes

$$\eqalign{{\cal R}_{\rho\alpha}&\biggl [{{\la\rho -1,n\ra}\over{\mu^2}}P^{\rho -1}_{\rho -1}
+\big\lbrace P^{\rho -1}_{\rho -1}+{{\la\rho -1,n-1\ra}\over{\mu^2}}P^{\rho -1}_{\rho -2}\big\rbrace (-1){{\la\beta ,n\ra}\over{\mu^2}}+\cr
+&\big\lbrace P^{\rho -1}_{\rho -2}+{{\la\rho -1,n-2\ra}\over{\mu^2}}P^{\rho -1}_{\rho -3}\big\rbrace
(-1)^2{{\la\beta ,n-1\ra\la\beta ,n\ra}\over{\mu^4}}+\dots\cr
\dots +&\big\lbrace P^{\rho -1}_{\rho -n+(\beta +1)}+{{\la\rho -1,\beta +1\ra}\over{\mu^2}}P^{\rho -1}_{\rho -n+\beta}\big\rbrace (-1)^{n-(\beta +1)}{{\la\beta ,\beta +2\ra ...\la\beta ,n\ra}\over{\mu^{2(n-(\beta +1))}}}+\cr
+&\big\lbrace P^{\rho -1}_{\rho -n+\beta}+{{\la\rho -1,\beta\ra}\over{\mu^2}}P^{\rho -1}_{\rho -n+\beta -1}\big\rbrace (-1)^{n-\beta}{{\la\beta ,\beta +1\ra ...\la\beta ,n\ra}\over{\mu^{2(n-\beta )}}}\biggr ]=\cr
={\cal R}_{\rho\alpha}&{{\la\rho -1,\beta\ra}\over{\mu^2}}\biggl [ P^{\rho -1}_{\rho -1}+P^{\rho -1}_{\rho -2}(-1){{\la\beta ,n\ra}\over{\mu^2}}+P^{\rho -1}_{\rho -3}(-1)^2{{\la\beta ,n-1\ra\la\beta ,n\ra}\over{\mu^4}}+\dots\cr
\dots +&P^{\rho -1}_{\rho -n+\beta}(-1)^{n-(\beta +1)}{{\la\beta ,\beta +2\ra ...\la\beta ,n\ra}\over{\mu^{2(n-(\beta +1))}}}+\cr
&+P^{\rho -1}_{\rho -n+\beta -1}(-1)^{n-\beta}{{\la\beta ,\beta +1\ra ...\la\beta ,n\ra}\over{\mu^{2(n-\beta )}}}\biggr ]\quad .}$$
\hfill (C.8)

The expression inside the brackets is the contribution to (C.1)
from the terms with $\rho -1$ divided by ${\cal R}_{\rho
-1,\alpha}\,$. So, if the contribution for $\rho$ is zero, it is
zero too for $\rho -1$ as long as $\la\rho -1,\beta\ra\ne 0$. When
$\beta <\alpha$, we have seen that for $\rho =n$ we have a null
contribution and then it is null for $\rho =n-1$, and so is down
to $\rho =\alpha +1$ and $\rho =\alpha$. All the contributions
being zero, the kinetic matrix is diagonal because of the
symmetry. For $\beta =\alpha$ the same procedure works well until
we reach $\rho =\alpha +2$, so that the contribution for $\rho
=\alpha +1$ is zero, but here the chain stops because $\la\rho
-1,\beta\ra =0$ in the next step, and we get

$$\eqalign{\bigl ({\cal R}^T\bar{\cal K}{\cal R}\bigr )_{\alpha\alpha}&={{\mu^{n(n-1)}}\over M}\sum\limits^{}_{\rho\ge\alpha }{\cal R}_{\alpha\alpha}P^{\alpha}_{\sigma -n+\alpha}{\cal R}_{\sigma\alpha}=\cr
&={\mu^{n(n-1)}\over{M}}(-1)^{n-\alpha}{{\la\alpha ,\alpha +1\ra ...\la\alpha ,n\ra}\over{\mu^{2(n-\alpha )}}}\biggl [P^{\alpha}_{\alpha}+(-1){{\la\alpha ,n\ra}\over{\mu^2}}P^{\alpha}_{\alpha -1}+\cr
&+(-1)^2{{\la\alpha ,n-1\ra\la\alpha ,n\ra}\over{\mu^4}}P^{\alpha}_{\alpha-2}+\dots\cr
&\dots +(-1)^{n-\alpha +1}{{\la\alpha ,\alpha +2\ra...\la\alpha ,n\ra}\over{\mu^{2(n-(\alpha +1))}}}P^{\alpha}_{\alpha -n+\alpha +1}+\cr
&+(-1)^{n-\alpha}{{\la\alpha ,\alpha +1\ra...\la\alpha ,n\ra}\over{\mu^{2(n-\alpha)}}}P^{\alpha}_{\alpha -n+\alpha}\biggr ]\quad.}$$
\hfill (C.9)

To get (4.30), we only need to prove that the square bracket in (C.9) gives

$${{\la1,\alpha\ra...\la\alpha -1,\alpha\ra}\over{\mu^{2(\alpha -1)}}}\quad.$$
\hfill (C.10)

To obtain this result, we apply to (C.9) the same procedure used in (C.8) and get

$${{\la\alpha -1,\alpha\ra}\over{\mu^2}}\biggl [P^{\alpha -1}_{\alpha -1}+(-1){{\la\alpha ,n\ra}\over{\mu^2}}P^{\alpha -1}_{\alpha -2}+\dots +(-1)^{n-\alpha}{{\la\alpha ,\alpha +1\ra...\la\alpha ,n\ra}\over{\mu^{2(n-\alpha)}}}P^{\alpha -1}_{\alpha -n+\alpha -1}\biggr ] .$$
\hfill (C.11)

We observe that $\alpha -n+\alpha\le\alpha$. If $\alpha -n+ \alpha
<0$ some of the terms in (C.9) are zero. If $\alpha -n+\alpha
=l\ge 0$ we can repeat for (C.11) the same operation we did on
(C.9), and, after $l+1$ steps, the last term is zero. After the
step $\alpha -2$ we finally get

$${{\la 2,\alpha\ra...\la\alpha -1,\alpha\ra}\over{\mu^{2(\alpha -2)}}}\biggl [P^2_2+(-1){{\la\alpha ,n\ra}\over{\mu^2}}P^2_1\biggr ]={{\la 1,\alpha\ra...\la\alpha -1,\alpha\ra}\over{\mu^{2(\alpha -1)}}}$$
\hfill (C.12)

\noindent{as} required.

Now we show how to get (4.31). We will study the situation when
$\beta\le\alpha$ as long as the mass matrix is symmetric, and
again omit ${\mu^{n(n-1)}\over M}\,$, so we consider

$$\sum\limits^{}_{\rho\ge\alpha ,\sigma\ge\beta}{\cal R}_{\rho\alpha}{\bar{\cal M}}_{\rho\sigma}{\cal R}_{\sigma\beta}\quad .$$
\hfill (C.13)

Take $\alpha =n$. Then, if $\beta =n$, one gets

$$-M^2_n\la 1,n\ra\la 2,n\ra...\la n-1,n\ra$$
\hfill (C.14)

\noindent{for} (C.13). If $\beta<n$ it is the same expression
obtained for the kinetic matrix with the factor $-M^2_n\,$, so it
vanishes.

For $\alpha =n-1$ we have in (C.13) $\rho =n,n-1$. The first value
gives the same result as in the kinetic case but with the factor
$-M^2_n\,$, and it vanishes again for $\beta\le\alpha$. For $\rho
=n-1$ we have contributions with and without $\mu^2$:

$$\eqalign{&\mu^2\biggl [P^n_{n-1}(-1){{\la\beta ,n\ra}\over{\mu^2}}+P^n_{n-2}(-1)^2{{\la\beta ,n-1\ra\la\beta ,n\ra}\over{\mu^4}}+\dots\cr
 &\dots +P^n_{\beta}(-1)^{n-\beta}{{\la\beta ,\beta +1\ra...\la\beta ,n\ra}\over{\mu^{2(n-\beta)}}}\biggr ]+(-M^2_n)P^{n-1}_{n-1}+\cr
&+(-M^2_{n-1})\biggl [P^{n-1}_{n-2}(-1){{\la\beta ,n\ra}\over{\mu^2}}+P^{n-1}_{n-3}(-1)^2{{\la\beta ,n-1\ra\la\beta ,n\ra}\over{\mu^4}}+\dots\cr
&\dots +P^{n-1}_{\beta -1}(-1)^{n-\beta}{{\la\beta ,\beta +1\ra...\la\beta ,n\ra}\over{\mu^{2(n-\beta)}}}\biggr ]\quad .}$$
\hfill (C.15)

Inside the bracket for $\mu^2$, we have the same expression
obtained for the kinetic part when $\rho =n$ but without the first
term, so it gives $-P^n_n\mu^2$. Adding it to the next, we have

$$-P^n_n\mu^2+P^{n-1}_{n-1}(-M^2_n)=-{{\la n-1,n\ra}\over{\mu^2}}P^{n-1}_{n-1}\mu^2+P^{n-1}_{n-1}(-M^2_n)=P^{n-1}_{n-1}(-M^2_{n-1})$$
\hfill (C.16)

\noindent{that}, together with the rest of (C.15), is the same
occurring in the kinetic case aside the factor $-M^2_{n-1}\,$. So
it is zero if $\beta<\alpha$, and for $\beta =\alpha =n-1$ one has
the result needed in (4.31).

Similar rearrangements are used for the rest of the matrix. For
example, for $\alpha =n-2$ one has two relations

$$\eqalign{&-P^{n-1}_{n-1}\mu^2+P^{n-2}_{n-2}(-M^2_n)=P^{n-2}_{n-2}(-M^2_{n-2})\cr
&-P^{n-1}_{n-2}\mu^2+P^{n-2}_{n-2}\mu^2+P^{n-2}_{n-3}(-M^2_{n-1})=\cr
&=-{{\la n-2,n-1\ra}\over{\mu^2}}P^{n-2}_{n-3}\mu^2+P^{n-2}_{n-3}(-M^2_{n-1})=P^{n-2}_{n-3}(-M^2_{n-2})\quad,}$$
\hfill (C.17)

\noindent{and} one arrives again at the kinetic result with the
factor $-M^2_{n-2}\,$.
\bigskip


\noindent{\bf Appendix D}
\bigskip

Consider the four-derivative Lagrangian,

$${\cal L}^{4} = -{1 \over 2}{1\over {\la 12\ra}}
              \,\phi \kg 1\gk\kg 2\gk \phi
                                          -j\,\phi\quad ,$$ \hfill (D.1)

\noindent{which} also reads

$${\cal L}^{4} = -{1 \over 2}{1 \over {\la 12\ra}}
       [\,p\,\phi^2 + s\,\phi(\square\,\phi)+ (\square\,\phi)^2]
                                          -j\,\phi\quad ,$$     \hfill (D.2)

\noindent{where} $p=m_1^2m_2^2$ and $s=m_1^2 + m_2^2\,$. The
covariant Ostrogradski method, in a slightly less refined version
that the one presented in [9], would define a conjugate
generalized momentum $\pi={{\partial{\cal
L}}\over{\partial(\square\,\phi)}}\,$. The Legendre transformation
performed on it leads to a Hamiltonian-like density from which the
following two-derivative Helmholtz Lagrangian is derived

$$ {\cal L}_H = \pi\,\square\,\phi + {1\over 2}\la 12\ra\,\pi^2
               + {1\over 2}\la 12\ra\,\phi^2 + {1\over 2}s\,\pi\phi\,.$$  \hfill (D.3)

\noindent{On} the other hand, by using the auxiliary field
technique of [17] the higher-derivative term is brought to second
order by writing (D.2) as

$${\cal L}^{4} = -{1 \over 2}{1 \over {\la 12\ra}}
                 [\,p\,\phi^2 + s\,\phi(\square\,\phi)
              + \Lambda(\square\,\phi) - {1\over 4}\Lambda^2]
                                          -j\,\phi\quad ,$$       \hfill   (D.4)

\noindent{where} the equation of motion for $\Lambda$ recovers
(D.2) when substituted back in (D.4). Now, in spite of their quite
different look, (D.3) and (D.4) are related by the simple field
redefinition $ \pi = -{1 \over 2}{1 \over {\la 12\ra}}(\Lambda +
s\phi)\,$.

\vfill
\eject


\centerline{\bf REFERENCES}\vskip 0.5cm

\noindent{$\,\,$} [1] K.Jansen, J.Kuti and Ch.Liu,
                                 {\it Phys.Lett.}{\bf B309}(1993)119;

                                 {\it Phys.Lett.}{\bf B309}(1993)127.

\noindent{$\,$} [2] B.Podolski and P.Schwed, {\it
Rev.Mod.Phys.}{\bf 20}(1948)40.

K.S.Stelle, {\it Gen.Rel.Grav.}{\bf 9}(1978)353.

\noindent{$\,$} [3] A.Bartoli and J.Julve, {\it Nucl.Phys.}{\bf
B425}(1994)277.

\noindent{$\,$} [4] K.S.Stelle, {\it Phys.Rev.}{\bf D16}(1977)953.

I.L.Buchbinder, S.D.Odintsov and I.L.Shapiro,

{\it Effective Action in Quantum Gravity},

\noindent{$\,$} [5] D.J.Gross and E.Witten,
                                 {\it Nucl.Phys.}{\bf B277}(1986)1.

R.R.Metsaev and A.A.Tseytlin,
                               {\it Phys.Lett.}{\bf B185}(1987)52 .

M.C.Bento and O.Bertolami,
                               {\it Phys.Lett.}{\bf B228}(1989)348.

\noindent{$\,$} [6] N.D.Birrell and P.C.W.Davies, {\it Quantum
Fields in Curved Space},

Cambridge Univ.Press(1982). (IOP, Bristol and Philadelphia, 1992).

\noindent{$\,$} [7] M.Ferraris and J.Kijowski, {\it Gen.Rel.Grav.}
                                                  {\bf 14}(1982)165.

A.Jakubiec and J.Kijowski, {\it Phys.Rev.}{\bf D37}(1988)1406.

G.Magnano, M.Ferraris and M.Francaviglia,
                                  {\it Gen.Rel.Grav.}{\bf 19}(1987)465;

                                  {\it J.Math.Phys.}{\bf 31}(1990)378;
                                  {\it Class.Quantum.Grav.}{\bf 7}(1990)557.

\noindent{$\,\,$} [8] J.C.Alonso, F.Barbero, J.Julve and
A.Tiemblo,

{\it Class.Quantum Grav.}{\bf 11}(1994)865.

\noindent{$\,$} [9] F.J.de Urries and J.Julve,
                                    {\it J.Phys.A: Math.Gen.}{\bf 31}(1998)6949.

\noindent [10] T.Nakamura and S.Hamamoto,
                                     {\it Prog.Theor.Phys.}{\bf 95}(1996)469.

\noindent [11] P.A.M.Dirac, {\it Canad.J.Math.}{\bf 2}(1950)129;
                          Proc.Roy.Soc.(London){\bf A246}(1958)326.
\noindent [12] E.Witten and \v C.Crnkovi\'c, {\it Covariant
description of canonical formalism in

geometrical theories}
 in {\it Three Hundred Years of Gravitation}
S.W.Hawking

and W.Israel Eds. Cambridge University Press (1989)
676-684.

J.Fernando Barbero G. and Eduardo J.S. Villase\~nor, {\it
Nucl.Phys.}{\bf B600}(2001)423

and references therein.

\noindent [13] F.Bopp, {\it Ann.Physik},{\bf 38}(1940)345.

B.Podolsky, {\it Phys.Rev.}{\bf 62}(1942)68.

B.Podolsky and C.Kikuchi, {\it Phys.Rev.}{\bf 65}(1944)228.

B.Podolsky and P.Schwed, {\it Rev.Mod.Phys.}{\bf 20}(1948)40.

\noindent [14] M.Kaku, {\it Phys.Rev.}{\bf D27}(1983)2819.

\noindent  [15]  A.Bartoli, J.Julve and  E.J.S\'anchez, {\it
Class.Quantum Grav.}{\bf 16}(1999)2283.

\noindent [16] M.Kaku, {\it Nucl.Phys.}{\bf B203}(1982)285.

\noindent [17] A.Hindawi, B.A.Ovrut and D.Waldram,
                                       {\it Phys.Rev.}{\bf D53}(1996)5583.

\vfill
\bye